# On the Existence of Synchrostates in Multichannel EEG Signals during Face-perception Tasks

Wasifa Jamal[a], Saptarshi Das[a], Koushik Maharatna[a], Fabio Apicella[b], Georgia Chronaki[c], Federico Sicca[b], David Cohen[d,e], Filippo Muratori[b]

**Authors' Affiliation:**

a) *School of Electronics and Computer Science, University of Southampton, Southampton SO17 1BJ, UK*
b) *IRCCS Stella Maris Foundation, Calambrone, Pisa, Italy*
c) *Section of Clinical & Cognitive Neuroscience, School of Psychological Sciences, University of Manchester, Manchester, UK*
d) *Groupe Hospitalier Pitié-Salpétrière, AP–HP, Paris, France*
e) *Institut des Systèmes Intelligents et de Robotiques, Université Pierre et Marie Curie, Paris, France*

**Authors' emails:**

wj4g08@ecs.soton.ac.uk (W. Jamal)

sd2a11@ecs.soton.ac.uk, s.das@soton.ac.uk (S. Das*)

km3@ecs.soton.ac.uk (K. Maharatna)

f.apicella@fsm.unipi.it (F.Apicella)

georgia.chronaki@manchester.ac.uk (G. Chronaki)

fsicca@fsm.unipi.it (F. Sicca)

david.cohen@psl.aphp.fr (D. Cohen)

filippo.muratori@fsm.unipi.it (F. Muratori)

**Phone:** +44(0)7448572598, **Fax:** 02380 593045

**Abstract:**

Phase synchronisation in multichannel EEG is known as the manifestation of functional brain connectivity. Traditional phase synchronisation studies are mostly based on time average synchrony measures hence do not preserve the temporal evolution of the phase difference. Here we propose a new method to show the existence of a small set of unique phase synchronised patterns or "states" in multi-channel EEG recordings, each "state" being stable of the order of ms, from typical and pathological subjects during face perception tasks. The proposed methodology bridges the concepts of EEG microstates and phase synchronisation in time and frequency domain respectively. The analysis is reported for four groups of children including typical, Autism Spectrum Disorder (ASD), low and high anxiety subjects – a total of 44 subjects. In all cases, we observe consistent existence of these states - termed as synchrostates - within specific cognition related frequency bands (beta and gamma bands), though the topographies of these synchrostates differ for different subject groups with different pathological conditions. The inter-synchrostate switching follows a well-defined sequence capturing the underlying inter-electrode phase relation dynamics in stimulus- and person-centric manner. Our study is motivated from the well-known EEG microstate exhibiting stable potential maps over the scalp. However, here we report a similar observation of quasi-stable phase synchronised states in multichannel EEG. The existence of the synchrostates coupled with their unique switching sequence characteristics could be considered as a potentially new field over contemporary EEG phase synchronisation studies.

## 1. Introduction

The intrinsic organisation of the human brain could be viewed as a dynamic network changing its configuration at sub-second level temporal scale depending upon a given cognitive task. Phase synchronisation dynamics between different cortical areas is fundamental to formulate a mathematical representation of such dynamically reconfiguring functional networks (Fell & Axmacher 2011), (Engel et al. 2001). Electroencephalography (EEG) is an effective tool for studying such phase synchronisation owing to its high temporal resolution and has been applied extensively in the past (Mulert et al. 2011)(Razavi et al. 2013), for studying such phenomena unearthing useful information about cognitive processes.

Traditionally, EEG based synchronisation analysis is mostly carried out at a time scale of the order of seconds, apart from the well-known microstate analysis (Thomas Koenig et al. 2002), where the scalp level distribution of electric field was studied at ms resolution level. Recently during a visual perception task it has been shown that at ms time scale there exists a small set of unique phase synchronised patterns, each being stable of the order of ms, and then abruptly switching from one to another (Wasifa Jamal et al. 2015). These quasi-stable phase difference patterns are termed as synchrostates. It was first observed in a single adult subject (Wasifa Jamal et al. 2013) and later in group average of typically developing as well as autistic children (W Jamal et al. 2013) in their respective EEG $\gamma$-bands. Subsequently functional connectivity networks were formulated from these $\gamma$-band synchrostates and then applying graph theoretic characterisation of them it was shown possible to classify autistic and typically developing population with high accuracy (Wasifa Jamal et al. 2014). This indicates towards the possibility of using synchrostates as a new way for functional connectivity network analysis in different populations. However the pertinent question is whether the synchrostates consistently exist in individual subjects in the cognition related bands ($\beta$ and $\gamma$) and if so, how much inter-subject variability could be expected with respect to the population average, as this is fundamental in ascertaining the possibility of classifying individual's pathological conditions using graph theoretic characterization of functional brain network formed from the synchrostates.

Therefore the main aims of this paper are: 1) systematically exploring the existence and nature of synchrostates in both the $\beta$ and $\gamma$ bands for individuals not only belonging to typically developing but also from pathological population, 2) studying variability of synchrostates with respect to the number of EEG electrodes used, 3) to elaborate the method of synchrostate formulation in step-by-step fashion showing how this method combines the existence of time-domain discrete state concept of microstates and frequency domain phase synchronisation analysis. The cognitive task selected for our exploration is a set of face-perception tasks where three types of face-perception related stimuli were given to four groups of children - with typical development, diagnosed with ASD, with high- and low-anxiety scores. Since $\beta$ and $\gamma$ bands have consistently shown increased synchrony during face-perception tasks (Uhlhaas et al. 2009)(Rodriguez et al. 1999)(Kottler et al. 2012) and prominent response during visual stimuli in general (Wróbel 2000)(Lachaux et al. 2005) we mainly concentrated on analysing these two bands. Our exploration showed: 1) the synchrostates exist in individual subjects consistently in both the $\beta$ and $\gamma$ bands and are usually bounded between 3 to 7 whereas in the low frequency bands ($\theta$, $\alpha$) there is no consistent existence of synchrostates; 2) synchrostates exhibit qualitatively similar behaviour as that of the EEG microstates in terms of their temporal stability and switching characteristics; 3) although the general set of synchrostate topographies are similar for a subject group corresponding to different visual perception stimuli, the actual time-courses of inter-synchrostate switching sequence are markedly different indicating towards stimulus-specific dynamics even within the broad category of visual perception task; and 4) using less number of electrodes results in greater variability in the number of synchrostates with respect to the corresponding population average whereas high density EEG gives more consistent result.

In addition, here we define all the synchrostates according to their topographical distribution of average phase difference over the scalp and reassign the class labels of similar topoplots with a state label which has been shown to vary little across different stimuli within the same population. We also observed that these synchrostates have different configurations in $\beta$ and $\gamma$ bands as well as across different subject groups. Also in order to quantify the qualitative behaviour of the synchrostate transitions, we calculated the probability of self-transition implying the degree of relative stability of these states in each combination of stimulus, population and frequency band.

## 2. Method
### 2.1 Background

Typically synchronisation can be studied from EEG signals in two domains i.e. time and frequency. The work reported in (D Lehmann et al. 1987)(Thomas Koenig et al. 2014) considered brain electric states with consistent scalp electric field topography and their sequence which lead to what is commonly known as EEG microstates. Its most important characteristic is that the topography does not change randomly or continuously over time but exhibit quasi-stable behaviour in the order of $80 - 120$ ms; and abruptly switches from one topography to another – the number of unique topographies being small (typically $3 - 10$) (Thomas Koenig et al. 2002). Another way to study the synchronisation phenomenon is in the frequency domain. This is led by the assumption that if two points (i.e. two EEG electrode sites) are in coherence (i.e. maintaining constant phase relationship over time), they can be considered as functionally synchronised or connected (Fries et al. 2001). Phase coupling has been studied in patients with mental disorders (Mulert et al. 2011)(Razavi et al. 2013) and the merits of synchrony analysis have been found in the understanding of neurodevelopmental disorders (Uhlhaas et al. 2008). Since the method for coherence analysis like Global Field Synchronisation (GFS) (Kottlow et al. 2012) use Fourier transform, inherently it does not preserve the temporal information of synchronisation. This methodology was later modified by several researchers by using Continuous Wavelet Transform (CWT) and Hilbert Transform (HT) to compute phase in transformed domains and for deriving associated synchronisation indices from the coherence values thus obtained. The mean phase coherence measure (Mormann et al. 2000) computes the synchronisation over the whole time series and therefore gives an average measure of synchronisation for the whole signal span. Phase Locking Value (PLV) although varies with time, measures the inter-trial variability of phase difference (Rodriguez et al. 1999) rather than temporal variability. Additionally, various other measures of phase synchronisation have been reported in (Quiroga et al. 2002). Although useful, such approaches only give insight into the phase synchronisation in a time-averaged way over all the frequency bands, rather than capturing the true picture of the temporal or transient evolution of phase synchrony in a band-specific way. On the other hand, in principle, CWT and HT both being time-frequency transform methods, have potential to describe the temporal evolution of phase synchronisation at sub-second resolution level which could be more informative to understand the dynamics of the synchronisation phenomena from the onset of a given stimulus till the end of the corresponding cognitive action.

As evident from the foregoing discussion the existing frequency domain methods compute the phase synchronisation over the entire post stimulus segment of the signal and therefore are unable to retain the transient information at finer temporal granularity, whereas the method of microstate finds the unique electric potential patterns and their transients during the execution period of the task (Gianotti et al. 2008). Ito *et al.* studied the dynamics of spontaneous transitions between globally phase-synchronised states in the alpha band EEG activity (Ito et al. 2007). Their method was applied to explore the phase dynamics of individuals with cerebral palsy (Daly et al. 2014). The technique proposed in these articles investigate phase dynamics by segmenting the relative phase patterns into global phase pattern states by thresholding and using a criterion called the Instantaneous Instability Index (III). The GPS pattern vectors are then clustered into 6 centroids. Here we propose a slightly different approach for studying the dynamical evolution of phase patterns by using intrinsic optimization criterion in the $k$-means clustering for segmentation of the images to form compact clusters or states without any prior thresholds. In this paper we merge two concepts, i.e. the concept of temporal switching (transient behaviour) of stable states along with the band specific phase locking by considering a joint time-frequency representation of the EEG signal.

### *2.2 Data and pre-processing*

The data analysis was conducted with four distinct samples of children: 1) with typical development, 2) diagnosed with Autism Spectrum Disorder (ASD), 3) diagnosed with high-anxiety and 4) with low-anxiety. More specifically, we have used the data acquired during the experiments described in (Fabio Apicella et al. 2013) and (Chronaki 2011). For more information regarding the data used in this study please refer to the supplementary material. The main characteristics of these four populations are summarised in Table 1.

Table 1: Summary of the subject group and presented stimuli

| Group number | Group type | Number of subjects | Age range | Number of EEG channels | Stimuli presented (types of faces) |
|---|---|---|---|---|---|
| I | Typical development | 12 | 6 - 13 | 128 | Happy, fear, neutral |
| II | ASD | 12 | 6 – 13 | 128 | Happy, fear, neutral |
| III | High-anxiety | 10 | 6 – 12 | 30 | Happy, angry, neutral |
| IV | Low-anxiety | 10 | 6 – 12 | 30 | Happy, angry, neutral |

For group I and II the data was acquired using 128 channel EEG system and was segmented into 1000 ms epoch with 150 ms baseline and 850 ms post-stimulus response. Epochs over a threshold of 200 µV were rejected as artefacts. Data was baseline corrected and band-pass filtered from $0.5 - 50$ Hz for removing low-frequency drifts and high-frequency measurement noise using $5^{th}$ order Butterworth filter. On the other hand, for group III and IV a 30 channel EEG system was used for data acquisition and data was epoched at 100 ms pre-stimulus to 1000 ms post-stimulus. Collected data was band-pass filtered in the range $0.1 - 70$ Hz for eliminating the drift and noise as done in the former case. These pre-processed EEG signals were then transformed in time-frequency domain using CWT and particularly focussing on the two bands of interest *viz.* $\beta$ (13-30 Hz) and $\gamma$ (30-50Hz). Only the response of the frequencies in these two bands of interest is then used for further processing. Being band-specific this processing step allows us to compare the response of the signals across all the groups on a uniform platform although they were acquired with different instruments. The studies in (Nunez et al. 1997)(Schiff 2005) point out that the use of average reference for calculating coherency is a good compromise against the effects and noise introduced by the reference electrode. In our study, all the signals were re-referenced to average reference data (average across all channels) and then were used for our calculation. The different data collection protocol adopted for group I - II; and III - IV allows us to explore the effect of variability in the number of EEG electrodes.

### 2.3 Computation of time-dependent phase difference topography

It has been observed by researchers that the spectral power of different EEG bands significantly changes depending upon the stimulus given (Boiten et al. 1992). As a consequence it may be assumed that the temporal stability of instantaneous phase difference topographies and hence the overall synchronisation pattern may manifest differently in different EEG bands. Therefore it appears to be more logical to study the synchronisation phenomenon in a band-specific way. Since CWT decomposes a signal to different scales (equivalent frequencies) at each time instant, it is possible to study the temporal evolution pattern of phase difference topographies for an isolated frequency band of interest. Therefore in our analysis we used CWT as the main analysis tool, more precisely, we have used a complex Morlet basis function as shown in (1) for computing the CWT of the EEG data.

$$\Psi_M(t) = \frac{1}{\sqrt{\pi F_b}} e^{2j\pi F_c t} e^{-\left(t^2/F_b\right)} \tag{1}$$

where, $\{F_b, F_c\}$ denote the bandwidth parameter and the centre frequency respectively. For our purpose we considered $F_b = 1$ and $F_c = 1.5$.

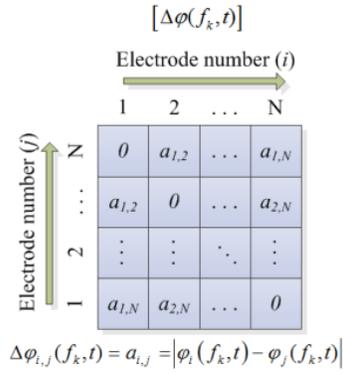

Figure 1: The structure of the phase difference matrix at frequency $f_k$ at time $t$.

Considering $N$ number of EEG channels placed over the scalp and $x_1(t), x_2(t), \cdots, x_N(t)$ be the EEG signals acquired at the respective channels, application of complex Morlet CWT on $x_i(t), i \in \{1, 2, \ldots, N\}$ results in a complex time series $W_i(a, t)$ at the wavelet scale $a$ at time $t$. $W_i(a, t)$ can be converted to a function of frequency and time $W_i(f, t)$ using the following relation (Addison 2010) in (2).

$$f = F_c / (a \cdot \delta) \tag{2}$$

where, $\delta$ and $f$ are the sampling period and the approximate pseudo-frequency, i.e. the frequencies corresponding to the scales, respectively. Subsequently, the instantaneous phase $\varphi_i(f, t)$ of $W_i(f, t)$ can be computed as (3).

$$\varphi_i(f, t) = \tan^{-1}\left(\frac{\text{Im}\left[W_i(f, t)\right]}{\text{Re}\left[W_i(f, t)\right]}\right) \tag{3}$$

$Im[W_i(f, t)]$ and $Re[W_i(f, t)]$ being the imaginary and the real part of $W_i(f, t)$ respectively. Consequently, the instantaneous phase difference $\Delta\varphi_{i,j}(f,t)$ between the channels $i$ and $j$ can be given by (4).

$$\Delta\varphi_{i,j}(f,t) = \left|\varphi_i(f,t) - \varphi_j(f,t)\right| \tag{4}$$

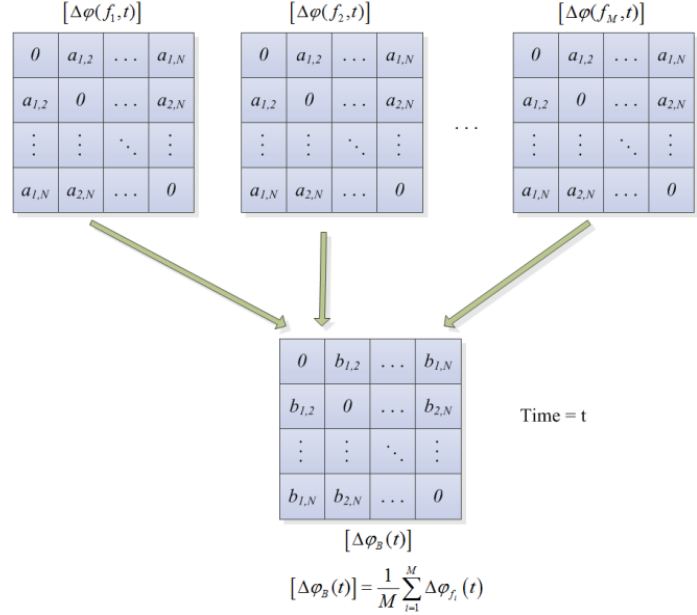

Figure 2: Computation principle of band-specific phase difference matrix.

Computation of $\Delta\varphi_{i,j}(f,t)$ at a time instant $t_1$ and frequency $f_1$ for $i, j \in \{1, 2, \ldots, N\}$ yields a symmetric square matrix $\left[\Delta\varphi(f_1,t_1)\right]$ that describes the pairwise relationship of phase difference at the frequency $f_1$ for all the EEG channels at $t_1$ time instant as shown in Figure 1. For computing the average response within a subject group the individual $\Delta\varphi_{i,j}(f,t)$ were averaged over all the subjects to get the average wavelet response for the group in consideration. Therefore, if the frequency band of interest $B \in \{\beta, \gamma\}$ is spanned over the frequencies $\{f_1, f_2, \cdots, f_M\}$ then the instantaneous phase difference matrix for $B$ at time $t$ as shown in Figure 2 can be formulated as (5)-(6).

$$\left[\Delta\varphi_B(t)\right] = \frac{1}{M}\sum_{i=1}^{M}\Delta\varphi(f_i,t) \tag{5}$$

$$(b_{i,j})_{\Delta\varphi_B}(t) = \frac{1}{M}\sum_{k=1}^{M}(a_{i,j})_{\Delta\varphi_{f_k}}(t) \tag{6}$$

where, $(b_{i,j})_{\Delta\varphi_B}(t)$ is the $(i, j)^{\text{th}}$ element of the matrix $\left[\Delta\varphi_B(t)\right]$ and $(a_{i,j})_{\Delta\varphi_{f_k}}(t)$ is the $(i, j)^{\text{th}}$ element of $\left[\Delta\varphi_{f_k}(t)\right]$. Subsequently, $\left[\Delta\varphi_B(t)\right]$ can be computed at different time instants $\{t_1, t_2, \cdots, t_n\}$ resulting in a set of such matrices $\left[\Delta\varphi_B(t_1)\right]$, $\left[\Delta\varphi_B(t_2)\right], \cdots, \left[\Delta\varphi_B(t_n)\right]$ like that shown in Figure 3 that describes the complete picture of temporal evolution of the phase difference from the onset of a stimulus till the end of the corresponding action in the particular frequency band $B$ over all the EEG channels on the scalp. The whole process is pictorially depicted in Figure 1-Figure 3. As evident, due the high dimensionality of the clustering problem, a simple cluster discrimination through a scatter plot is not possible in the present study. In our clustering framework, at each time instant the dimension of the feature space, denoting all possible cross-electrode phase information, becomes $N^2$ and the clustering algorithm considers the evolution of each element $b_{i,j}(t)$ along time.

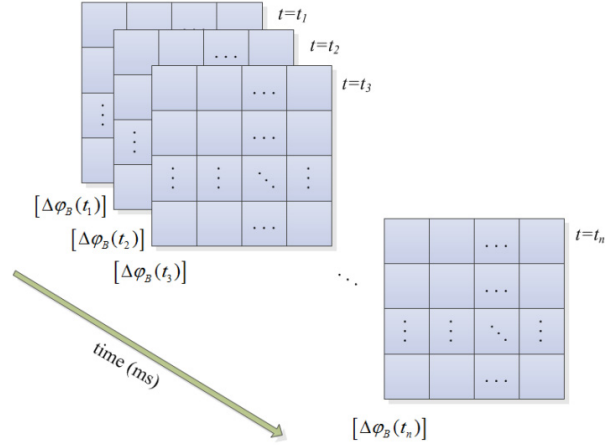

Figure 3: Computation of band-specific phase difference matrix from the onset of a stimulus till the end of the desired time window.

### 2.4 Clustering of phase difference matrices into unique set of 'states'

Once all the cross-electrode phase difference matrices for a particular band are formulated over the entire duration of a specified time interval – in our case, we are interested to see the temporal evolution of these topographies at subsecond order time interval – the next pertinent question is whether there exists any unique 'spatio-temporal pattern' of phase difference topographies during the execution of the cognitive task. The first step for that is to identify all 'possibly unique' topographies over the entire time duration of interest. A certain class of pattern recognition techniques could be employed for this purpose. The $k$-means clustering is one such unsupervised pattern recognition technique. For a give dataset $X$, $X = \{x_p\}, p \in [1, \cdots, P]$, assuming the number of underlying clusters is known, $k$-means algorithm iteratively minimises a cost function given below (7).

$$J(\theta, U) = \sum_{p=1}^{P}\sum_{q=1}^{m} u_{pq} \left\| x_p - \theta_p \right\|^2 \qquad (7)$$

where, $\theta = \left[\theta_{T1} \ldots \theta_{Tm}\right]^T$, $\|\cdot\|$ is the Euclidean distance, $\theta_q$ is the centre of a cluster and $u_{pq} = 1$ if $x_p$ lies closest to $\theta_q$; 0 otherwise. Initially $k$ centroids are defined and depending on how near the data vector is to the centroids, it is assigned to a class. The $k$ centroids are iteratively recalculated and the data is reassigned to these new centroids until the data vectors from $X$ form compact clusters and the cost function $J$ is minimised. Initially a range $[m_{min}, m_{max}]$ is defined for possible clusters $m$ for the dataset $X$. The $k$-means clustering runs $n$ ($n$ random initializations) times for each $m$ within that range and for every $n$ runs the minimum value of the cost function $J_m$ (as shown in (7)) is calculated and stored. The cost function $J_m$ essentially indicates the sum of distances of the data-points from the nearest cluster mean when $m$ clusters are considered. The value of $J_m$ is dependent on the number of clusters and also the dataset under consideration whereas a high value of $J_m$ represents a less compact cluster. Thus we search for a 'knee' in the plot of $J_m$ against $m$ as an indication of the number of optimal clusters underlying the data. If the plot of $J_m$ against $m$ shows a significant 'knee' at $m = m_1$ (say) then it signifies that the number of optimum clusters underlying the dataset $X$ is likely to be $m_1$. To be noted that in the plot of $J_m$ versus $m$ it is typical to have multiple such knees as $m$ varies within its selected range. In cases, where there is an increase in the $J_m$ value, it indicates that the distance between all the data points with respect to the nearest mean of clusters has increased. This increase could be due to the splitting of large compact clusters into several smaller ones, caused by increasing the value of $m$. In such a case, one need to consider the earliest and the most prominent knee as the *characteristic knee* and the corresponding $m$ as the underlying number of clusters as it explains the dataset with minimum complexity. Another important point to note that the absolute value of $J_m$ in the plot of $J_m$ against $m$ is not important but the value of $m$ at which $J_m$ attains minimum value (the significant knee) is the important parameter indicating the number of underlying clusters. This method is also known as incremental $k$-means or elbow method and is widely used to find the optimum number of clusters in a given data-set.

In a higher dimensional feature space, the landscape of the cost function $J(\theta)$ may have multiple local minima and there is small probability of finding a higher value of the cost function if a local minima has been found by the optimization process. Since the $k$-means clustering have the problem of getting trapped to local minima, it should be run multiple times with different initialization of the cluster means and the best result with the minimum value of the cost function should be considered. Therefore in our method, the best-results of the $k$-means algorithm for each choice of $k$ is considered out of $n = 10$ different random initializations of the cluster means. This way the incremental $k$-means plots the best cost function to obtain the $J_m$ as also suggested in (Theodoridis et al. 2010).

The unsupervised learning technique adopted here is based on the concept of hard clustering, i.e. a single data-point corresponding to each time instant should belong to one of the clusters, because in a temporal resolution of millisecond, we assume that the brain stays only in one state. Other paradigms of soft clustering like fuzzy $c$-means or similar methods (Dimitriadis et al. 2013) where a single data-point can be associated with more than one cluster according to its degree of associativity with different classes, could also be applied to the present problem.

In our case, $X$ is the dataset of all pairwise EEG instantaneous phase differences $\left[\Delta\varphi_B(t_1)\right], \left[\Delta\varphi_B(t_2)\right], ..., \left[\Delta\varphi_B(t_n)\right]$, as a function of time. We clustered the dataset $X$ along time $t$, for a chosen frequency band $B$, to find out unique phase difference patterns. The algorithm yields $k$ centroids, $\theta_j$ for each cluster or state and a vector of length $n$ with the corresponding state or cluster labels for each $\left[\Delta\varphi_B(t)\right]$ for every time instance over which we clustered. The centroids hold average information for each of the clustered states whereas the cluster labels signify when in time each state has occurred. Once the phase-difference matrices are uniquely clustered over different time instances, the centroids are translated into corresponding colour-coded head-map topographies following arbitrary colour coding convention. This is done by first calculating the average phase difference seen at a particular electrode with respect to the rest of the electrodes i.e. taking row-wise average of $[\Delta\varphi_B(t)]$ and considering it as the average phase difference at that electrode index corresponding to the considered row and assigning a particular colour corresponding to the numerical value of that phase difference and finally transforming it to a contour plot. Such head-map topographies give a visual representation of the distribution of average phase differences between different regions of brain over the scalp. Note that these plots should not be viewed or compared to the typical EEG potential plots or the power spectrum plot typically generated in quantitative EEG (qEEG) analysis. Here the plots show the gross phase difference between different electrodes over the scalp over a particular time window. Higher numerical values represent greater gross phase difference of the electrode with all the other electrodes and low values indicate that the electrode has relatively less phase difference with all the other electrodes in that configuration. We term the set of topography clusters identified using $k$-means algorithm as synchrostates. The state labels are used to construct a transition plot to illustrate the switching sequence of the synchrostates over the time of the EEG recording. This is simply done by plotting the time labels yielded by the clustering algorithm.

## 3. Results

As mentioned previously, we restrict our study only in the $\beta$ and $\gamma$ band since research indicated that they are directly related to the cognitive task related to face perception (Wróbel 2000)(Lachaux et al. 2005). We present the results in two steps: first as a population average and then for individual subjects belonging to a population. To study the population average we first formulate the average phase difference matrix for each subject by taking the mean of the phase difference matrices across all trails. Then we take an average of the phase matrices of each subject belonging to that population at every time instant and invoke $k$-means clustering on that set of average matrices as described in Section 2.2 and 2.3. In essence this gives a general picture of temporal evolution of phase relationship between different electrode sites for a specific population. Our exploration shows that the cost function for clustering does not fall arbitrarily with the increase in the number of clusters confirming the existence of a finite number of compact underlying clusters or states during the whole time-course of the EEG data. The detailed results for the individual groups are furnished in the following subsections.

### 3.1 Typical development

Figure 4 shows the results of the $k$-means clustering algorithm in both the bands for all the given stimuli from the population average of 12 children with typical development (group I). It is clear that the dominant knee of cost function for all the three stimuli appears at $k = 3$ although in some cases after the knee the cost function increases and then again decreases. These are the typical situations already discussed in Section 2.4 and accordingly where the earliest knee appeared needs to be considered only. This means that in the dataset considered, there exist three unique phase difference matrix configurations – synchrostates – from the onset of stimulus till the end of an action.

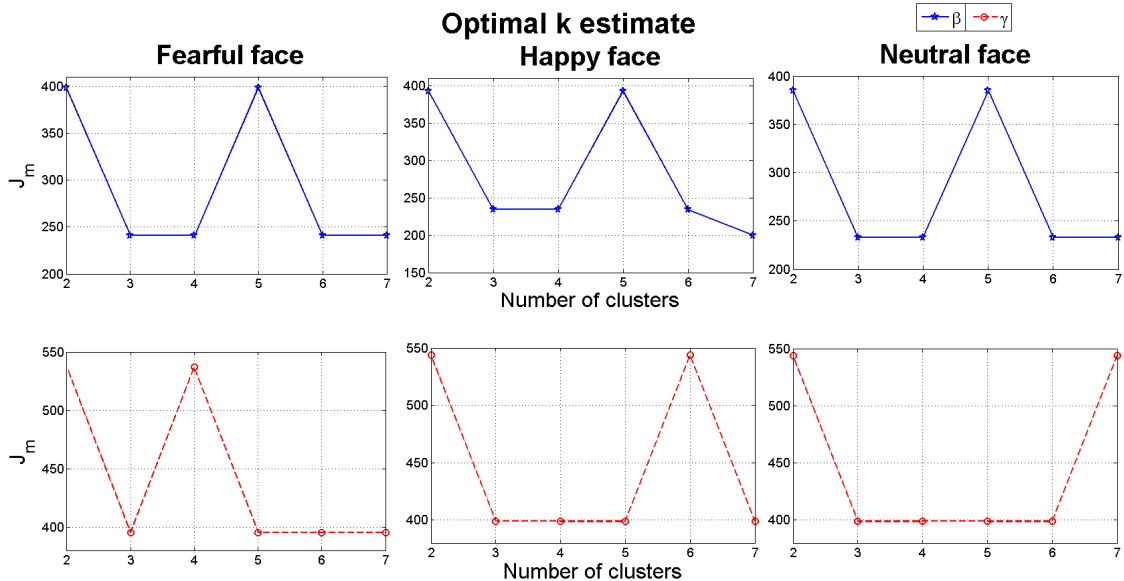

Figure 4: $k$-means clustering result of $\beta$ and $\gamma$ band for the typical group.

In Figure 5, from the corresponding head-plots it is evident that the topographies of all the three synchrostates are very similar for all the different stimuli in the $\beta$ band. A similar result is observed for the $\gamma$ band in Figure 6 where the synchrostate topographic plots are similar and more importantly closely resembles to those obtained in the $\beta$ band although differing slightly in the numerical values, in particular in the reddish hue regions. Here, each of the colours in the head topographies signify a particular range of phase differences as shown in the legend (in a normalized scale with respect to the maximum and minimum phase difference amongst all the states). However, an interesting difference is observed in the state transition plots shown in Figure 7. Although in both the bands the transitions start from state 2, the overall transition patterns are markedly different not only between the $\beta$ and $\gamma$ band but also between different stimuli within a band. This demonstrates the stimulus specific nature of the synchrostates.

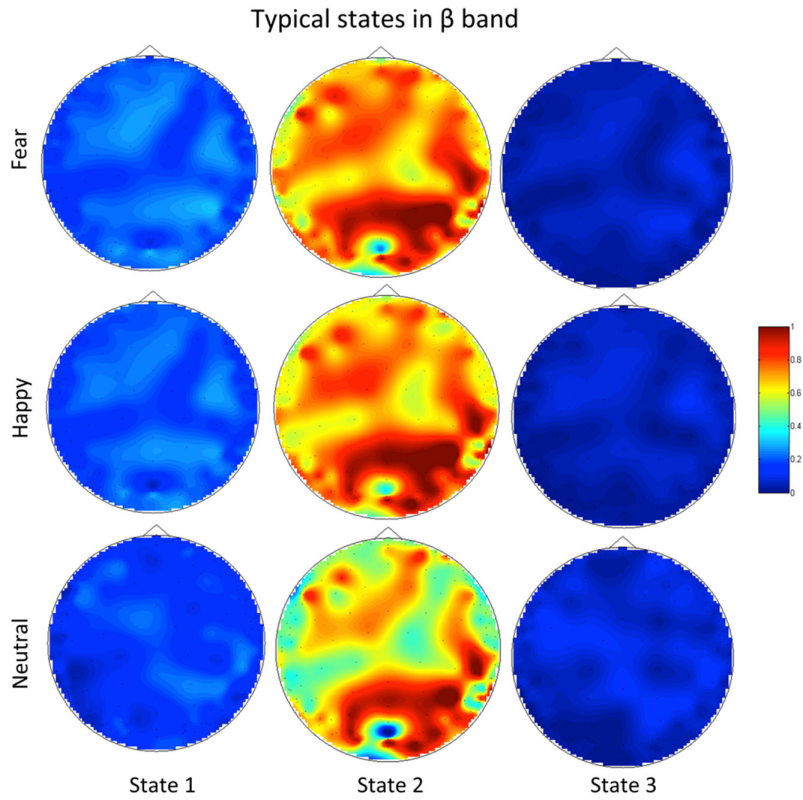

Figure 5: The topographic map for all the three stimuli in $\beta$ band for the typical group.

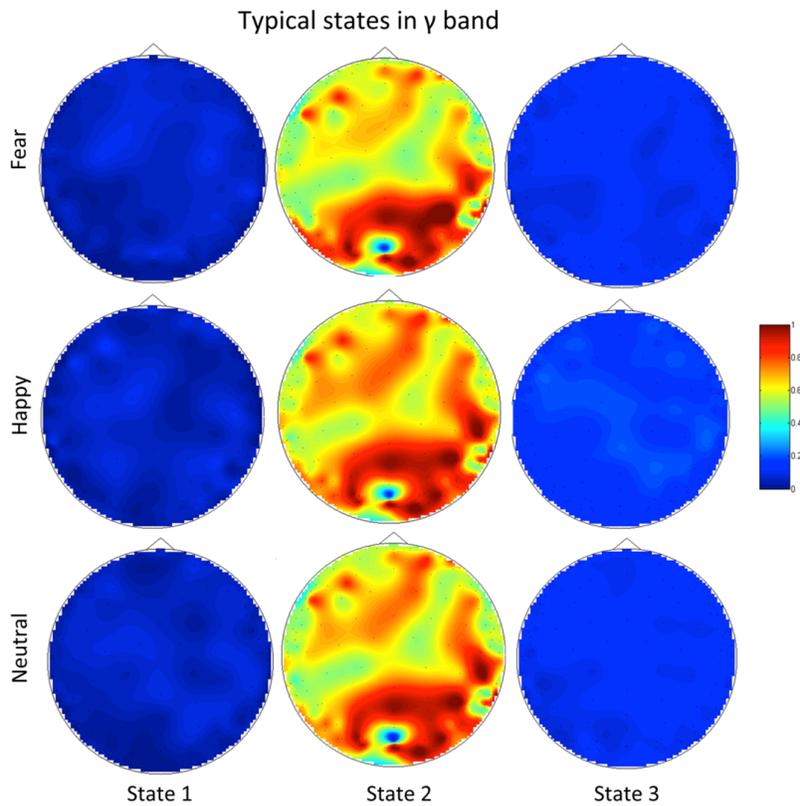

Figure 6: The topographic map for all the three stimuli in $\gamma$ band for the typical group.

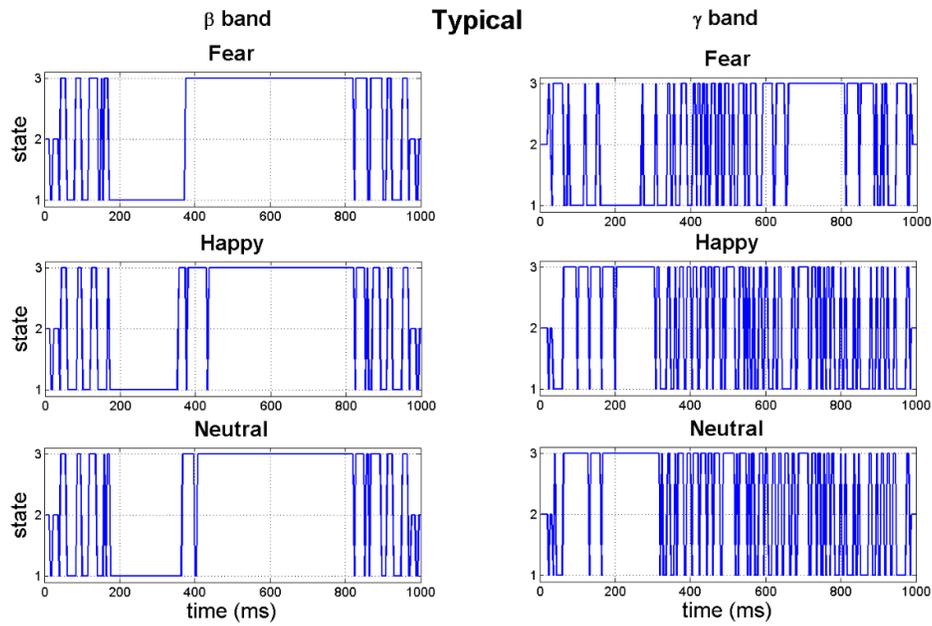

Figure 7: The time-course plot of synchrostate transitions in $\beta$ and $\gamma$ band.

### 3.2 ASD

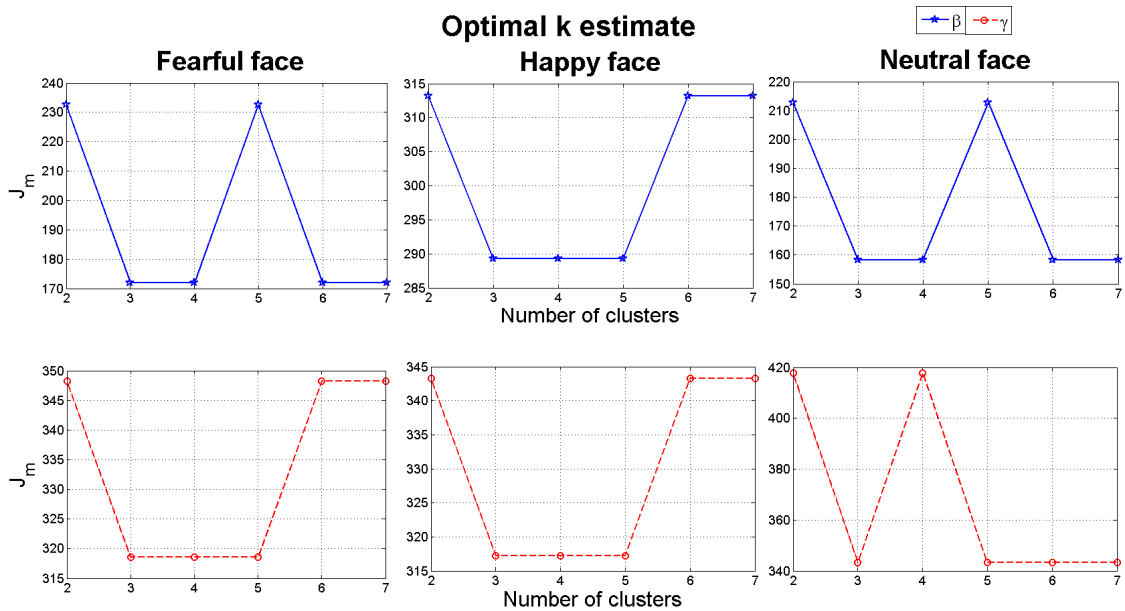

Figure 8: $k$-means clustering result of $\beta$ and $\gamma$ band for the ASD group.

The $k$-means clustering results for ASD population (Group II in Table 1) is shown in Figure 8 for the $\beta$ and $\gamma$ bands for all the three applied stimuli i.e. fearful, happy and neutral faces. Once again the significant knee appears at $k = 3$ implying existence of three synchrostates similar to the typical case. The corresponding phase difference topographies over the scalp are shown in Figure 9-Figure 10 as head plots. It appears that although the stimuli are different the topographies are nearly similar in the $\beta$ bands (in Figure 9) in particular for state 1 and state 3. However topographies corresponding to state 2 are markedly different. On the other hand, in $\gamma$ band the state 1 for happy and neutral stimuli are similar while it differs significantly for fear stimulus (Figure 10). State 3 shows close similarity under all the three stimuli. The time-course plots of the synchrostate transition are shown in Figure 11. In both of the bands the time course plots are markedly different depending upon the stimulus and thereby indicating different temporal stability period of the synchrostates at different points in time.

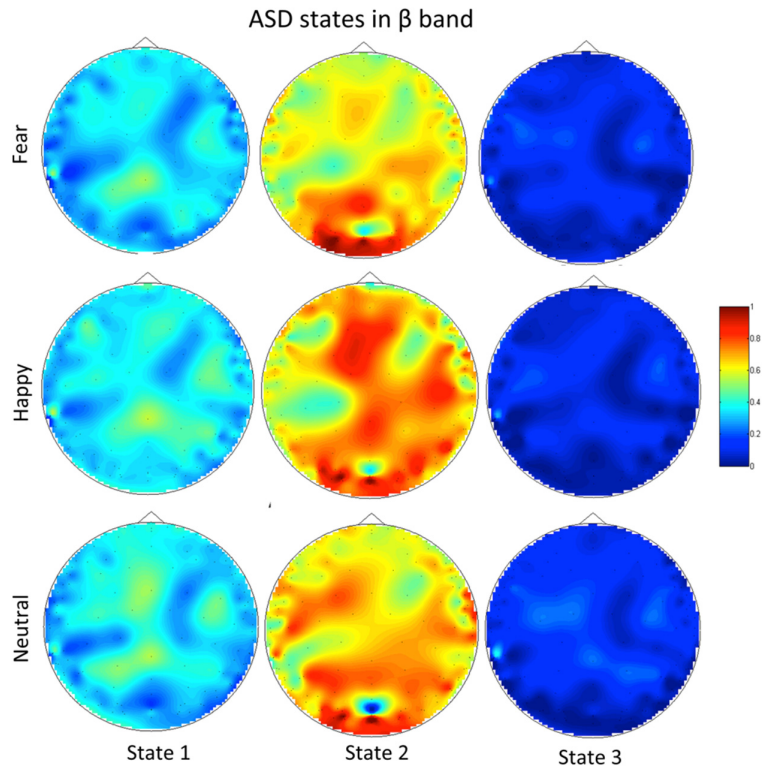

Figure 9: The topographic map all the three stimuli in $\beta$ band for the ASD group.

ASD states in γ band

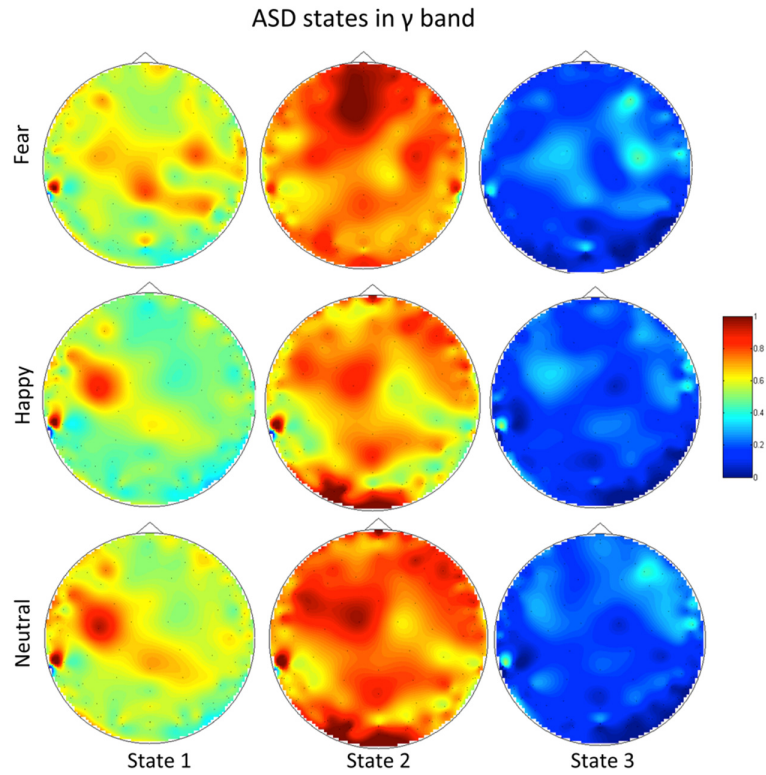

Figure 10: The topographic map for all the three stimuli in γ band for the ASD group.

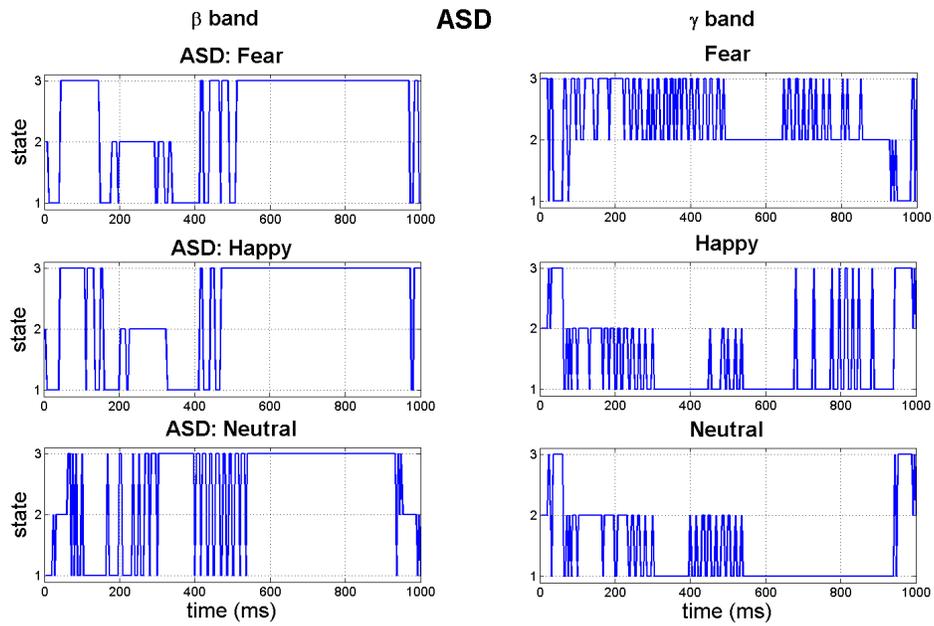

Figure 11: The time-course plot of synchrostate transitions in β and γ band.

### 3.3 Low Anxiety

The *k*-means clustering when run on the population average of the children with low anxiety for the *β* band resulted with four states for all the three stimuli i.e. angry, happy and neutral face. This is shown in Figure 12 as all three plots have the earliest significant 'knee' in the cost function plot at *k* = 4. However in the *γ* band the number of states is different for the neutral face perception case. The number of states in the *γ* band for angry and happy face remain unchanged at *k* = 4 whereas for neutral face it is 6. In the head topographies for the *β* band, (Figure 13) although the number of synchrostate is consistently four, their characteristics for each different task are quite different. From the *γ* band head plots in Figure 14 it can be seen that the states 1, 4, 5 and 6 head plots are almost similar and common for all the three stimuli. However the neutral stimulus has two extra states which do not exist in the other two stimuli of angry and happy as can be seen from Figure 14. The transition of the states in *β* band is shown for each specific stimulus in Figure 15. It can be observed that during the execution of the angry face stimulus the inter-state transition is not as frequent as compared to the other two stimuli *viz.* happy and neutral. The *β* band state transition shows that except for angry stimulus for both the other stimuli the sequence start with state 4, whereas for the angry visual stimulus it starts from state 2. In the *γ* band, the state transitions are more frequent in neutral face perception compared to the other two stimuli as shown in Figure 15.

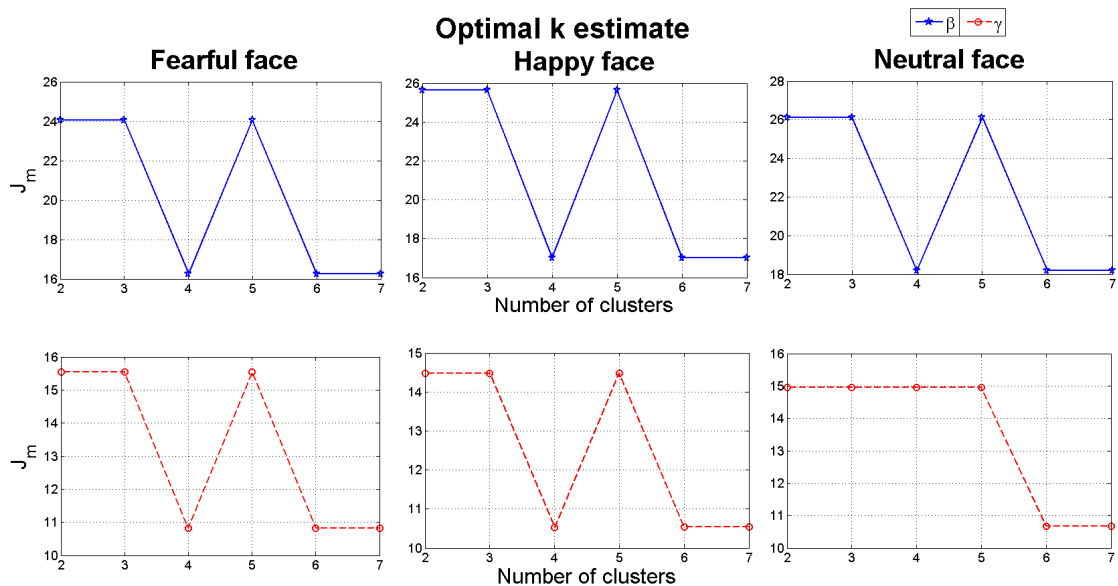

Figure 12: *k*-means clustering result of *β* and *γ* band for the low anxiety group.

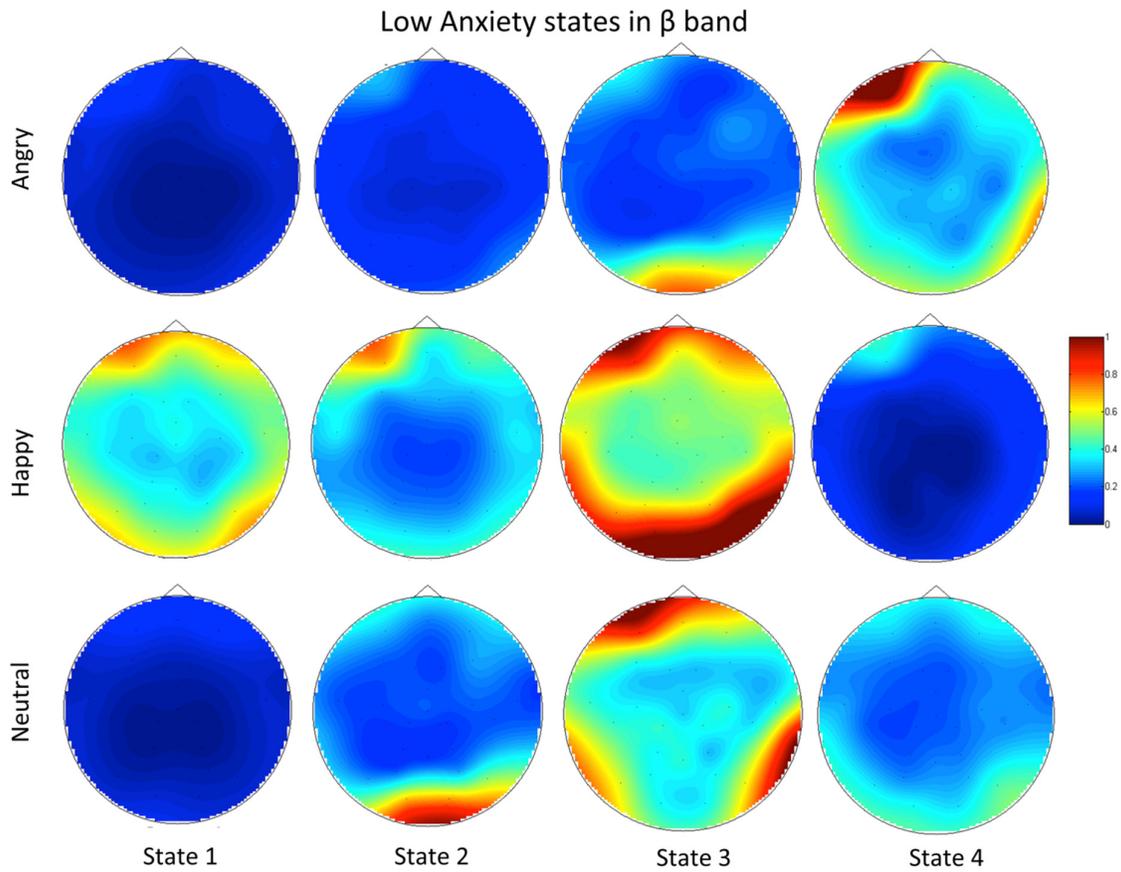

Figure 13: The topographic map for all the three stimuli in *β* band for the low anxiety group.

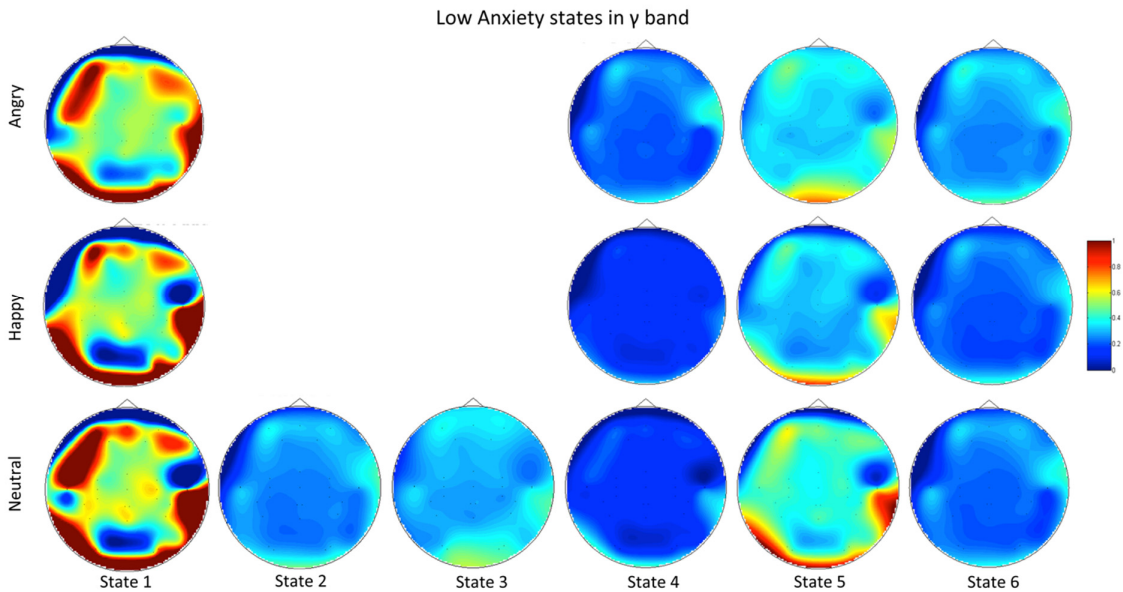

Figure 14: The topographic map for all the three stimuli in *γ* band for the low anxiety group.

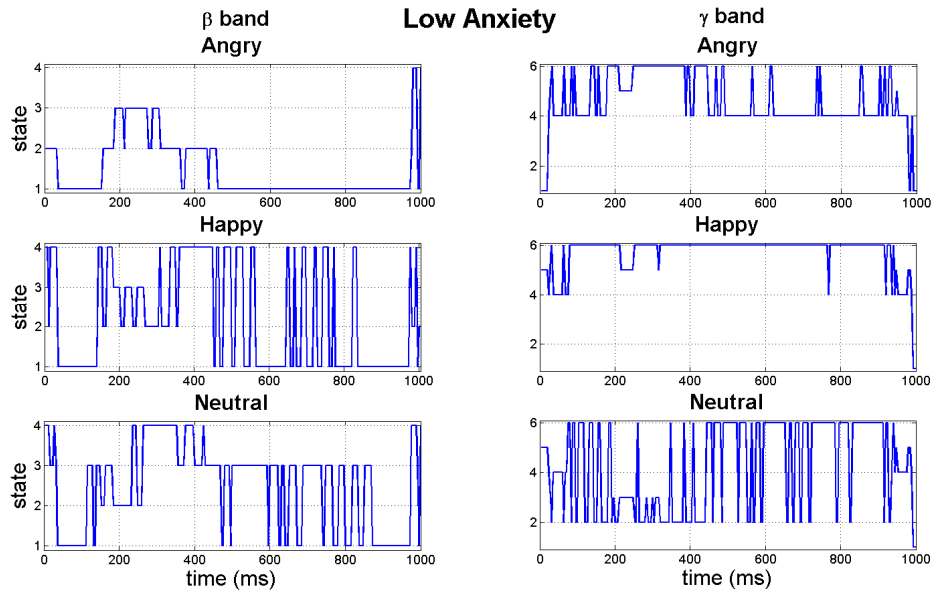

Figure 15: The time-course plot of synchrostate transitions in $\beta$ and $\gamma$ band.

### 3.4 High Anxiety

For the group of high anxiety subjects, as shown in Figure 16, for both the bands the number of synchrostates is consistently four for different stimuli. The head plots for the average $\beta$ responses of the children as can be seen from Figure 17 are to some extent similar across all the stimuli. This close similarity is even more prominent in the $\gamma$ band head plots depicted in Figure 18. Looking at the transitions of the states in $\beta$, shown in Figure 19 we see that they end in state 1 for all the three stimuli. Also state 3 is the most occurring state over the duration shown for happy and neutral face. This is also the case for $\gamma$ band state transitions for all stimuli as shown in the Figure 19.

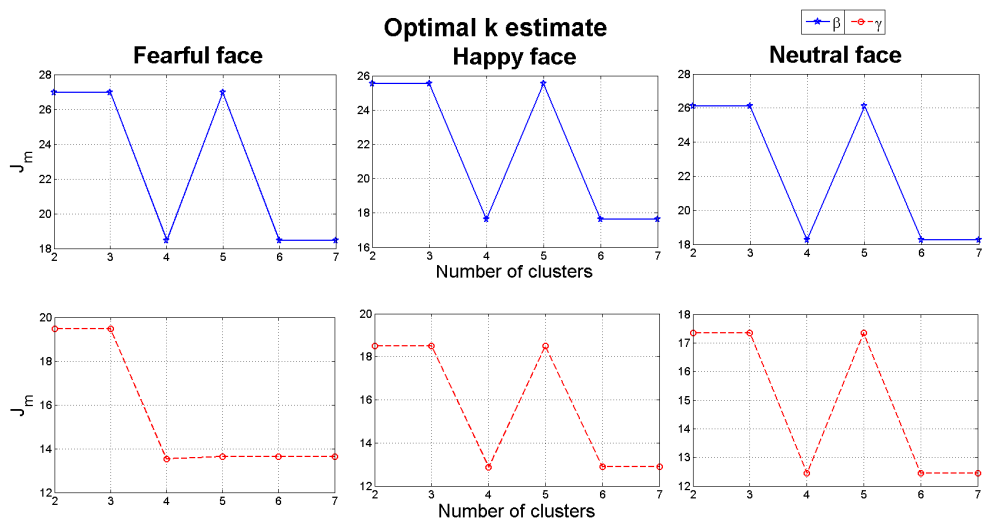

Figure 16: $k$-means clustering result of $\beta$ and $\gamma$ band for the high anxiety group.

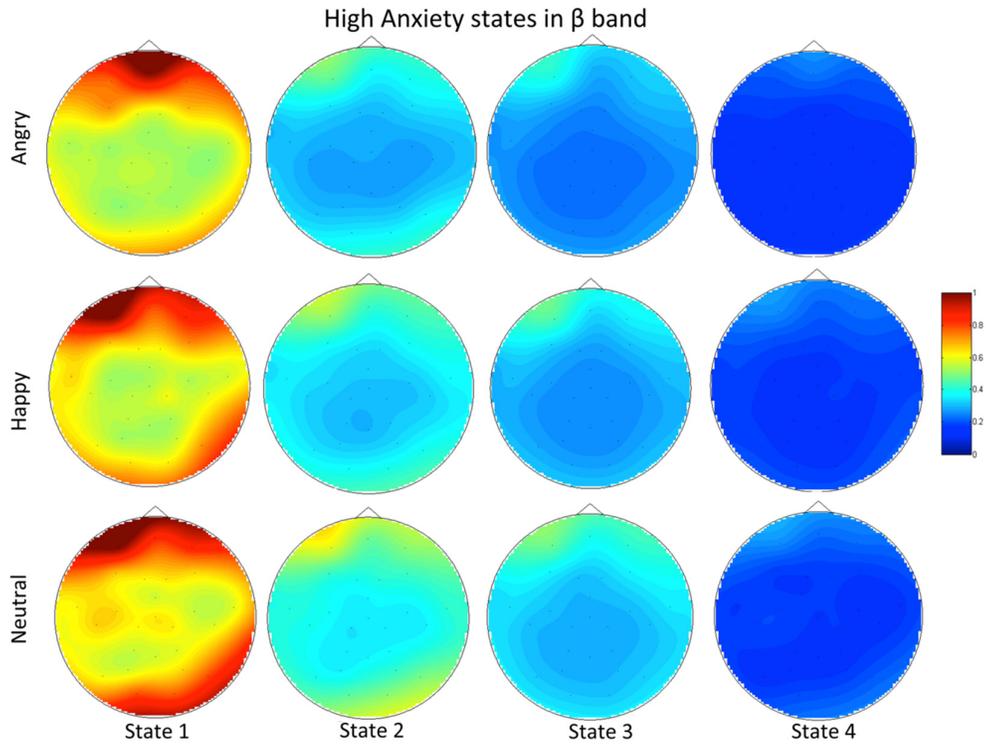

Figure 17: The topographic map for all the three stimuli in $\beta$ band for the high anxiety group.

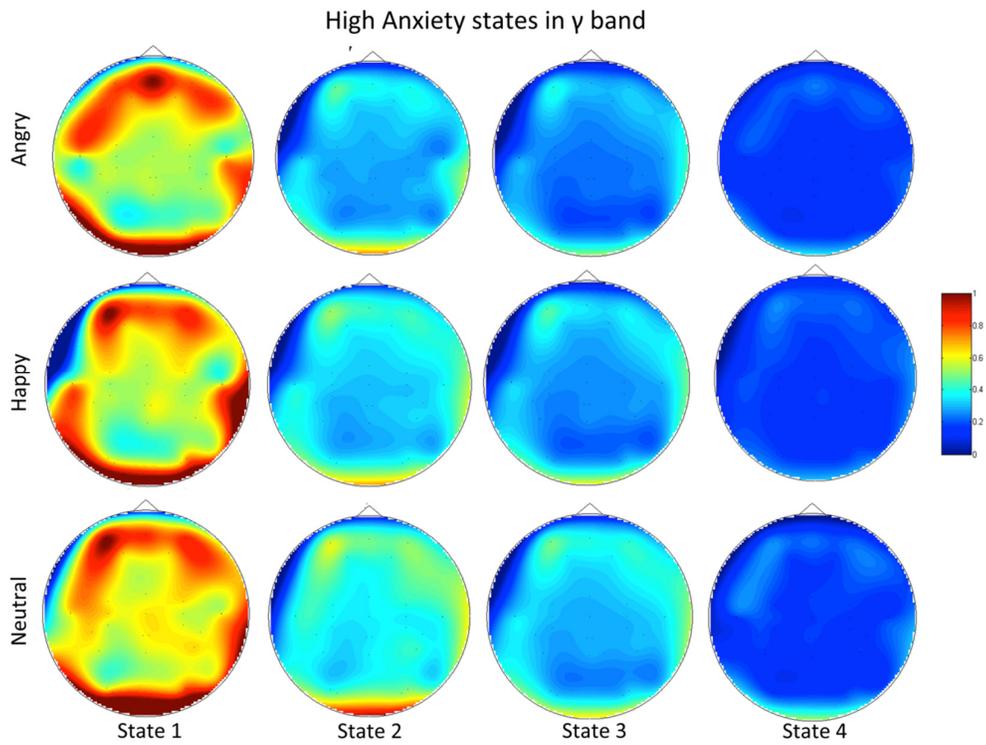

Figure 18: The topographic map for all the three stimuli in $\gamma$ band for the high anxiety group.

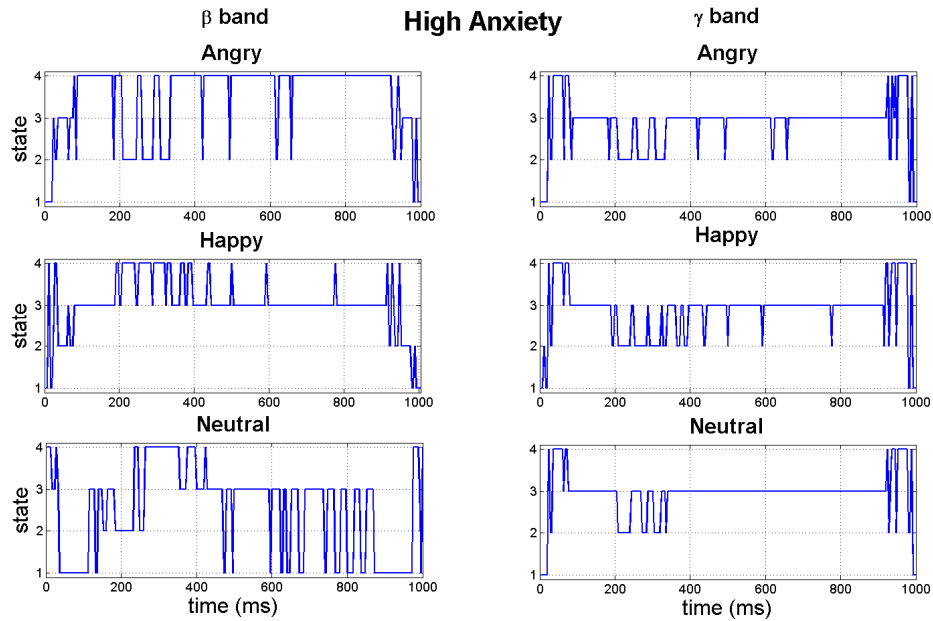

Figure 19: The time-course plot of synchrostate transitions in $\beta$ and $\gamma$ band.

### 3.5 Variability analysis for individual subjects

So far the reported figures for the group-wise analysis highlight subtle changes in the average phase difference topographies over the scalp and state transition plots for different stimuli. Now the statistical measures like the median, inter-quartile ranges of the inter-person variability for the optimal number of synchrostates in both $\beta$ and $\gamma$ band are shown in the box-plots given in Figure 20. The red line in the plot indicates the median and the crosses show the outliers. The blue boxes denote the inter quartile range for the data. This is obtained by applying $k$-means clustering on the phase-difference matrices obtained from individual subjects at different time instants under different stimuli. The variability in the number of synchrostates observed when results from the individual subjects are compared to the respective population average is not significant. For the pool of typical children we got consistently three states for every child in the $\gamma$ band but in the $\beta$ band the number of states for the children varies from 3-7. This observation leads us to believe that the number of synchrostates is person-specific although this number is bounded within a small range only. Also in Figure 20, for ASD group in the $\beta$ and $\gamma$ band only few subjects show 5 synchrostates whereas the population average result as well as for the other subjects, the number of synchrostates is consistently 3. For the low anxiety and high anxiety groups (low-density EEG) it is interesting to note that the median of the number of synchrostates varies between 5 and 6 whereas the median is consistently 3 for the ASD and typical children (high-density EEG). The important factor to note here is that only 30 electrodes were used for EEG acquisition for the anxiety groups (III and IV). This reduced number of electrodes inherently introduced less resolution in computing the phase difference matrix and as a consequence may introduce a larger variability in the synchrostate formulation. Therefore it is evident that the optimal number of synchrostates largely depends on the number of electrodes and high-density EEGs (as in the first two groups, Typical and ASD) are more likely to give consistent result. Apart from that, the small variability observed in all the four cases is also expected because of inter-person and inter-trial variability and possible existence of parallel background processes not related to the cognitive task given.

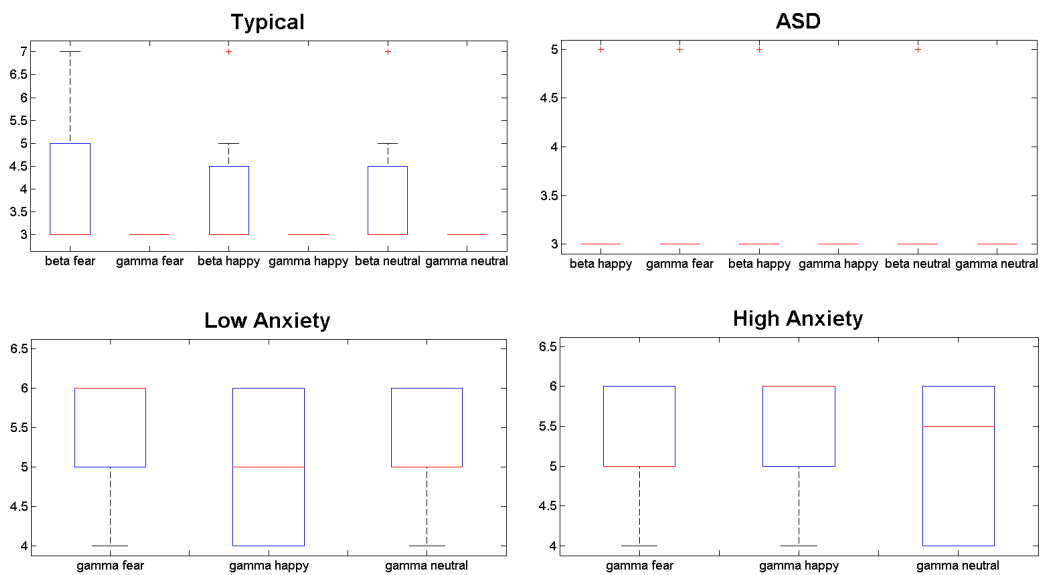

Figure 20: Box-plot of the variation in the optimal number of synchrostates in each group of subjects.

### 3.6 Quantification of the synchrostate transition

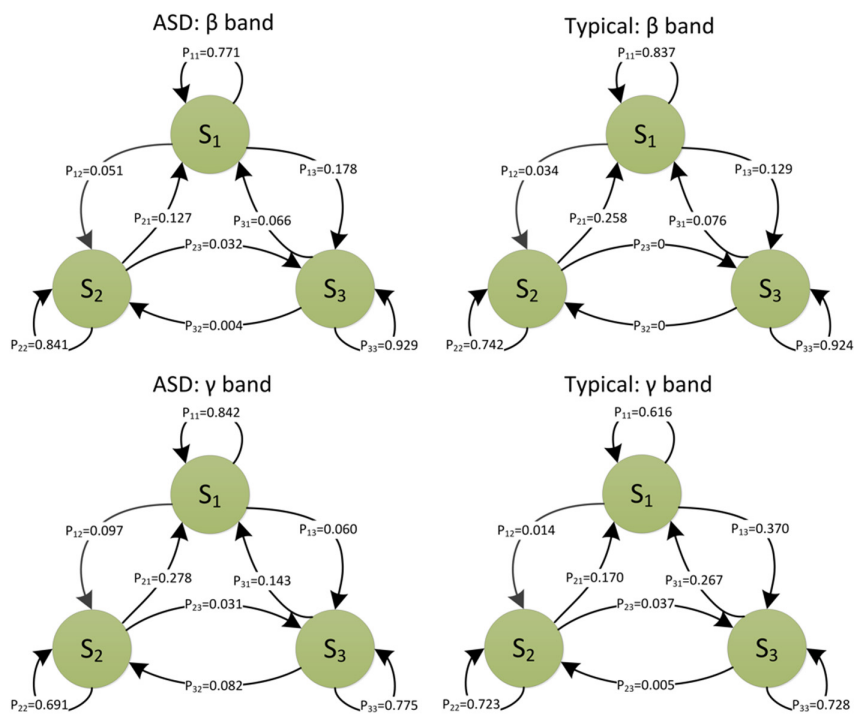

Figure 21: Average state transition (across different stimuli) diagrams for the typical and ASD group in the $\beta$ and $\gamma$ band

We now model the temporal switching sequence of the synchrostates in a probabilistic framework for a representative case of the typical and ASD group. This is chosen due to the fact that the high-density EEG system used in data acquisition for these two groups resulted into consistent results in synchrostates as mentioned earlier. However without loss of generality the same method could be applied for analyzing the anxiety groups as well. We construct the transition probability ($P_{ij} = n_{ij} / \sum_j n_{ij}$) of the synchrostate sequence which show the probabilistic nature of each of the state transitions. Here, $n_{ij}$ is the number of transitions from state $i$ to $j$. The three probability values - $P_{11}$, $P_{22}$, and $P_{33}$ show how long each state remain stable i.e. how stable each of the states ($S_1, S_2, S_3$) are in terms of the probability of staying in the same state, for different population groups, as shown in Figure 21. The elements of the state transition matrix ($P_{ij}, \{i, j\} \in [1, 2, 3]$) for different population are more informative, although the phase difference topographies for two different populations could be similar. Therefore, the average value of the self-transitions ($(1/N)\sum_{i=1}^{N} P_{ii}$, $N$ being the optimal number of synchrostates) for a particular band, can be considered as one of the discriminating measure between two groups as shown in Table 2. It is evident from the Table 2 that in the $\beta$ band with fear and happy stimuli the ASD group has got a higher probability of self-transition than the typical case. On contrary the $\gamma$ band shows an increase in self-transition for fear and neutral stimuli.

Table 2: Self-transitions in $\beta$ and $\gamma$ band for the typical and ASD group with different stimuli

| Stimuli | Typical | | ASD | |
|---|---|---|---|---|
| | $\beta$ | $\gamma$ | $\beta$ | $\gamma$ |
| Fear | 0.843293 | 0.756196 | 0.880439 | 0.880439 |
| Happy | 0.834324 | 0.695356 | 0.904088 | 0.674856 |
| Neutral | 0.880439 | 0.684694 | 0.757696 | 0.752797 |

## 4. Discussion

The results shown in the foregoing section indicate that when observing in the sub-second order temporal resolution the phase difference topography and hence phase synchronisation between EEG electrodes distributed over the scalp for all the four populations is bounded within a small number (typically 3 to 7) unique patterns. These results are based on the sensor or scalp level EEG synchronisation analysis. It is well known that often the scalp level synchronization with zero phase lag could be confounded by the effect of volume conduction. Therefore, a similar method for synchrony analysis could be done at the source level. Because of the lack of spatial resolution in EEG, source level synchrony gives more reliable physiological interpretations. But the very nature of source level synchrony analysis suffers from the lack of temporal resolution restricting such methods for carrying out transient analysis of brain dynamics at fine temporal granularity level. One way to capture this effect is to translate the EEG to the corresponding source level using an inverse mapping techniques. However it is well recognized that reconstructing source activity from EEG is an ill-posed problem and using EEG alone cannot uniquely determine the spatial locations of the underlying sources. Theoretically, only an infinite number of electrodes on the scalp would allow the unique determination of the locations of the sources inside (Koles 1998). Therefore, one has to make some assumptions about the inverse problem, to obtain optimal and unique solution which leads to approximate sources (Phillips et al. 2005).

### 4.1 Possible artifact and volume conduction effect

Before continuing discussion about the implications of this result one needs to eliminate possible artefact effect that may bias the observation. Once again we need to emphasise that the head plots shown here are fundamentally different from those obtained from qEEG analysis where the average power spectrum is plotted over the scalp. Any possible artefact in such cases is manifested as strong correlation at the scalp edges. On the contrary, the head plots shown here are more like the visualisation of the phase difference patterns distributed over the scalp. The bluish hues imply nearly zero phase difference whereas the reddish hue implies large phase difference.

Here, each of the head plots show phase difference topographies existing of the order of ms. While processing the data, as mentioned in Section 2, we eliminated the epochs above 200 μV as possible artefacts. Therefore the data used in our analysis is likely to be artefact free in the first place. Secondly, since the synchrostate topographies are constructed in the ms order and as the transition diagrams show that the topographies switch from one configuration to another and back, in the ms order time interval, in the presence of possible artefacts all of the states should exhibit similar phase relation at the scalp edge for all the states which is not the case. Therefore while interpreting the results one may eliminate the effect of possible artefacts. This argument is also valid for eliminating the effect of volume conduction which one may also view as possible artefacts. The synchrostate phenomenon we report here cannot be explained by volume conduction since electrical impulses within the human brain spread almost instantaneously through any volume. Hence, zero phase lag is a characteristic of volume conduction (Thatcher et al. 2008) and phase delays are attributes of network formation. Synchrostates do not report zero phase delays and thus suggests the synchronies are not artefactual. In addition, the spurious synchronisation phenomena typically observed due to the presence of volume conduction does not account for the different synchronisation and desynchronisation patterns in the ms order in the switching characteristics of the synchrostates as such effect is expected to be present for all the synchrostate topographies in that case while in reality the synchrostate topographies for a stimulus are different from each other. Synchrony caused due to volume conduction would render a constant synchronisation pattern (phase difference) throughout the scalp over the observed time-course of the signal. This is not the case in synchrostates as the phase patterns change suddenly both in strength and between electrodes over time and again remains stable for a finite duration. From these points one may conclude that the results shown in this work are not due to possible artefacts but manifestation of the phenomena of transient phase difference dynamics triggered by different face perception stimuli.

### 4.2 Existence of synchrostates

The most important finding of this study is that over the four different subject groups and a total of 44 subjects a small set of unique phase difference patterns – synchrostates – each being stable of the order of ms have been found to exist in the β and γ band. These synchrostates switch from one to another abruptly and thereby constructing a characteristic time-course to the applied stimulus. This is qualitatively similar to the results obtained with microstates (T Koenig et al. 2005) albeit the microstate topographies are constructed in the EEG amplitude domain where the number of states is up to 10. From our experiments, we observe that the number of synchrostates is bounded between 3 and 7 depending on individual subjects, stimuli and also the number of EEG electrodes for recording. From the time-course plots it is evident that different synchrostates show different duration of stability at different time point depending on the applied stimulus and thereby possibly capturing the dynamics of phase synchronisation at a finer temporal granularity level.

An interesting observation for the case of typical development population is that the topographies of population average synchrostates are almost similar for all the applied stimuli in both the bands. This is similar to our earlier observation (Wasifa Jamal et al. 2013) where initial exploration was carried out with single normal adult subject. Intuitively this implies that although different stimuli have been applied, since all of them belong to the general class of face perception task, the fundamental phase relationship over the scalp remains nearly the same indicating a specific type of information integration phenomenon pertaining to the general face perception scenario. However the effect of different stimuli within the general class of face perception is reflected in the respective time-course plot which showed marked difference as the characteristic of the applied stimulus. On the other hand, although the topographies in the case of ASD population showed certain similarities they are more variable compared to the typical case along with their time-course. This may be due to the difference in information processing in the brain between the two subject groups. In addition, it is apparent that generally for the ASD group the gross phase difference of each electrode across the scalp is higher than that compared to the typical group as there is more presence of red and yellow hues in the ASD states in the $\gamma$ band (Figure 10) compared to the more blue hues in the states for the typical (Figure 6) group. This implies predominantly loose synchronisation in the former case which also falls in the line of already established theory that ASD brains show less synchronisation in information processing compared to the typically growing children. Individuals with autism present atypical neural activity in face processing and eye gaze tasks and this has been associated with later diagnosed autism (Elsabbagh et al. 2012). Similar considerations apply to the children with anxiety. However as discussed in Section 3.5 it seems that determination of the optimal number of synchrostates depends on the electrode systems used for EEG recording and more consistent result could be obtained using high-density system. Given this fact a direct comparison between group I-II and III-IV could be misleading as they do not share the same number of electrode configurations. But despite this fact it is evident that the number of synchrostates in all the four cases does not vary widely and is bounded within a small number of 3 to 7 depending on the pathophysiological conditions of the subject.

### 4.3 Physical Interpretation

Synchrostates are the states within which the inter-electrode relative phase difference varies little over time and the corresponding transition plot indicates how each of these phase-difference topographies remain stable. Hence the interpretation of these states cannot be done in an isolated way from its transition plots. Interpretation of the synchrostate topographies and the state transitions should be done together combining the stability duration and their respective numerical values of phase difference. When considered together one can formulate a synchronisation index corresponding to each of the synchrostates from which scalp-level functional connectivity network could be derived. These dynamic networks are governed by the nature of switching patterns of the synchrostates and therefore in essence capture the temporal evolution of functional connectivity in stimulus-specific way at fine temporal granularity level. Fundamental graph-theoretic measures could be used for characterizing such networks for gaining quantitatively deeper insight into the temporal dynamics of the connectivity pattern prevailing after the onset of stimuli and therefore may provide a quantitative means for assessing cognitive functionalities. This approach has been adopted in (Jamal et al. 2014) to classify a population of typical and ASD children. Therefore, synchrostates and their associated temporal switching sequences may be considered as a new tool for analyzing the scalp-level functional connectivity dynamics at a fine temporal scale given a type of stimulus indicating towards the dynamics of information exchange in a person-centric, and frequency band-specific manner. A major implication is that comparing the graph-theoretic measures, extracted from the functional connectivity networks formulated through synchrostates may provide a new way of classifying different neurodevelopment disorders.

Why such synchrostates exist and, given the fact that EEG has poor spatial resolution, how the phase difference described by the synchrostates corroborate with the actual anatomical level (or source level) connectivity and information exchange, is still an open question and requires further experiments and modelling activities. Thus the neurological perspective of synchrostate topographies, their numbers and transitions needs to be explored in future research. Another important fact is that the results reported here are only for face perception tasks. Whether the same phenomenon exists with other types of stimuli, e.g. auditory stimuli or different real-life cognitive activities is still a question to answer. Also whether the existence of synchrostates is associated only with the active cognitive states or not, is an area to explore.

## 5. Conclusions

Our analysis described in this paper shows that there exist a small set of unique phase difference patterns at ms order time interval amongst the EEG electrodes when 44 subjects from three different neuro-pathological groups and one healthy group were subjected to a set of facial perception task. These unique patterns – termed as synchrostates – abruptly switch from one to another and construct a stimulus-specific time course. The synchrostates and their transition plots can together be utilized as a generic method to understand temporal dynamics of EEG phase synchronisation as was done in (Jamal *et al.* 2014). Our present exploration shows that existence of such synchrostates is consistent and exhibits only a small variability that may be attributable to inter-person or inter-trial variation often expected to be present in such experiments. Another possible factor that may contribute in such variability is the number of electrodes – less number of EEG electrodes exhibiting greater variability by introducing less resolution in computing the phase difference pattern. Also quantification of the synchrostate transition in different groups are done in a probabilistic frame in terms of the self-transitions which might help in understanding the EEG phase synchronisation based derivation of the functional brain connectivity. Although we observed consistent number of synchrostates their physiological origin in relation to the anatomical brain network is yet to be established. Also it is still an open question whether the existence of synchrostates is a general phenomenon associated with active cognitive computation. However if established as a generic phenomenon, combining the phase topographies of the synchrostates and their temporal stability from the time-course plot, one may establish a set of quantitative index that may give deeper understanding in transient phase relationship with effective connectivity in brain which may be useful in quantifying cognitive ability in task-specific manner as well as classifying atypical neuropsychiatric conditions from normal brain functionality.

**Acknowledgement**

The work presented in this paper was supported by FP7 EU funded MICHELANGELO project, Grant Agreement # 288241. URL: www.michelangelo-project.eu/.

**References**

Addison, P.S., 2010. *The illustrated wavelet transform handbook: introductory theory and applications in science, engineering, medicine and finance*, CRC Press.

Apicella, Fabio et al., 2013. Fusiform Gyrus responses to neutral and emotional faces in children with Autism Spectrum Disorders: a High Density ERP study. *Behavioural Brain Research*, 251, pp.115–162.

Boiten, F., Sergeant, J. & Geuze, R., 1992. Event-related desynchronization: the effects of energetic and computational demands. *Electroencephalography and Clinical Neurophysiology*, 82(4), pp.302–309.

Chronaki, G., 2011. *A behavioural and electrophysiological exploration into facial and vocal emotion processing in children with behaviour problems*.

Daly, I. et al., 2014. Exploration of the neural correlates of cerebral palsy for sensorimotor BCI control. *Frontiers in neuroengineering*, 7.

Dimitriadis, S., Laskaris, N. & Tzelepi, A., 2013. On the Quantization of Time-Varying Phase Synchrony Patterns into Distinct Functional Connectivity Microstates (FCμstates) in a Multi-trial Visual ERP Paradigm. *Brain Topography*, 26(3), pp.397–409.

Elsabbagh, M. et al., 2012. Infant neural sensitivity to dynamic eye gaze is associated with later emerging autism. *Current Biology*, 22(4), pp.338–342.

Engel, A.K., Fries, P. & Singer, W., 2001. Dynamic predictions: oscillations and synchrony in top-down processing. *Nature Reviews Neuroscience*, 2(10), pp.704–716.

Fell, J. & Axmacher, N., 2011. The role of phase synchronization in memory processes. *Nature Reviews Neuroscience*, 12(2), pp.105–118.

Fries, P. et al., 2001. Modulation of oscillatory neuronal synchronization by selective visual attention. *Science*, 291(5508), pp.1560–1563.

Gianotti, L.R. et al., 2008. First valence, then arousal: the temporal dynamics of brain electric activity evoked by emotional stimuli. *Brain Topography*, 20(3), pp.143–156.

Ito, J., Nikolaev, A.R. & Leeuwen, C. van, 2007. Dynamics of spontaneous transitions between global brain states. *Human brain mapping*, 28(9), pp.904–913.

Jamal, W et al., 2013. Using brain connectivity measure of EEG synchrostates for discriminating typical and Autism Spectrum Disorder. In *Neural Engineering (NER), 2013 6th International IEEE/EMBS Conference on*. pp. 1402–1405.

Jamal, Wasifa et al., 2015. Brain connectivity analysis from EEG signals using stable phase-synchronized states during face perception tasks. *Physica A: Statistical Mechanics and its Applications*, 434, pp.273–295.

Jamal, Wasifa, Das, Saptarshi & Maharatna, Koushik, 2013. Existence of millisecond-order stable states in time-varying phase synchronization measure in EEG signals. In *Engineering in Medicine and Biology Society (EMBC), 2013 35th Annual International*

*Conference of the IEEE*. pp. 2539–2542.

Jamal, Wasifa et al., 2014. Classification of autism spectrum disorder using supervised learning of brain connectivity measures extracted from synchrostates. *Journal of Neural Engineering*, 11(4), p.046019.

Koenig, T et al., 2005. Brain connectivity at different time-scales measured with EEG. *Philosophical Transactions of the Royal Society B: Biological Sciences*, 360(1457), pp.1015–1024.

Koenig, Thomas et al., 2014. A Tutorial on Data-Driven Methods for Statistically Assessing ERP Topographies. *Brain Topography*, 27(1), pp.72–83.

Koenig, Thomas et al., 2002. Millisecond by millisecond, year by year: normative EEG microstates and developmental stages. *Neuroimage*, 16(1), pp.41–48.

Koles, Z.J., 1998. Trends in EEG source localization. *Electroencephalography and Clinical Neurophysiology*, 106(2), pp.127–137.

Kottlow, M. et al., 2012. Increased phase synchronization during continuous face integration measured simultaneously with EEG and fMRI. *Clinical Neurophysiology*, 123(8), pp.1536–1548.

Lachaux, J.-P. et al., 2005. The many faces of the gamma band response to complex visual stimuli. *NeuroImage*, 25(2), pp.491–501.

Lehmann, D, Ozaki, H. & Pal, I., 1987. EEG alpha map series: brain micro-states by space-oriented adaptive segmentation. *Electroencephalography and Clinical Neurophysiology*, 67(3), pp.271–288.

Mormann, F. et al., 2000. Mean phase coherence as a measure for phase synchronization and its application to the EEG of epilepsy patients. *Physica D: Nonlinear Phenomena*, 144(3), pp.358–369.

Mulert, C. et al., 2011. Long-range synchrony of gamma oscillations and auditory hallucination symptoms in schizophrenia. *International Journal of Psychophysiology*, 79(1), pp.55–63.

Nunez, P.L. et al., 1997. EEG coherency: I: statistics, reference electrode, volume conduction, Laplacians, cortical imaging, and interpretation at multiple scales. *Electroencephalography and Clinical Neurophysiology*, 103(5), pp.499–515.

Phillips, C. et al., 2005. An empirical Bayesian solution to the source reconstruction problem in EEG. *NeuroImage*, 24(4), pp.997–1011.

Quiroga, R.Q. et al., 2002. Performance of different synchronization measures in real data: a case study on electroencephalographic signals. *Physical Review E*, 65(4), p.041903.

Razavi, N. et al., 2013. Shifted coupling of EEG driving frequencies and FMRI resting state networks in schizophrenia spectrum disorders. *PloS One*, 8(10), p.e76604.

Rodriguez, E. et al., 1999. Perception's shadow: long-distance synchronization of human brain activity. *Nature*, 397(6718), pp.430–433.

Schiff, S.J., 2005. Dangerous phase. *Neuroinformatics*, 3(4), pp.315–317.

Thatcher, R.W., North, D.M. & Biver, C.J., 2008. Development of cortical connections as measured by EEG coherence and phase delays. *Human Brain Mapping*, 29(12), pp.1400–1415.

Theodoridis, S. et al., 2010. *Introduction to Pattern Recognition: A Matlab Approach: A Matlab Approach*, Academic Press.

Uhlhaas, P.J. et al., 2009. The development of neural synchrony reflects late maturation and restructuring of functional networks in humans. *Proceedings of the National Academy of Sciences*, 106(24), pp.9866–9871.

Uhlhaas, P.J. et al., 2008. The role of oscillations and synchrony in cortical networks and their putative relevance for the pathophysiology of schizophrenia. *Schizophrenia Bulletin*, 34(5), pp.927–943.

Wróbel, A., 2000. Beta activity: a carrier for visual attention. *Acta Neurobiologiae Experimentalis*, 60(2), pp.247–260.

*__Supplementary Material__*

Standard EEG acquisition protocol was followed for acquiring the data at 250 samples/s. The children in group I were typically healthy children with a mean age of 9.7 years. The subjects for group II were diagnosed with ASD according to the Diagnostic and Statistical Manual of Mental Disorders (DSM-IV-TP) criteria (Association 2000). The diagnosis was confirmed by Autism Diagnostic Observation Schedule-Generic (ADOS-G) and Autism Diagnostic Interview-Revised (ADI-R). The stimuli for the experiment run with group I and II were taken from a database of widely used standardized facial expressions (Tottenham et al. 2009). 30 faces from 5 male and 5 female subjects were taken, each exhibiting fearful, happy and neutral expressions. The subjects in group III and IV are children with anxiety problems and were assessed using the DOMINIC (Valla et al. 2000) which is a DSM-IV based pictorial interview for children aged 6-11 years. Children in this group met the recommended cut-off points for generalized anxiety (Valla et al. 2000). Group I and IV should not be confused although the level of symptoms in the low anxiety group might be similar to that of typically developing children since the later have been only assessed for generalized anxiety. The pictures for the stimuli for these two groups were standardized pictures of angry, happy and neutral facial expressions taken from FEEST database (Young et al. 2002). The example stimuli for group I and II; and III and IV are shown in Figure 1.

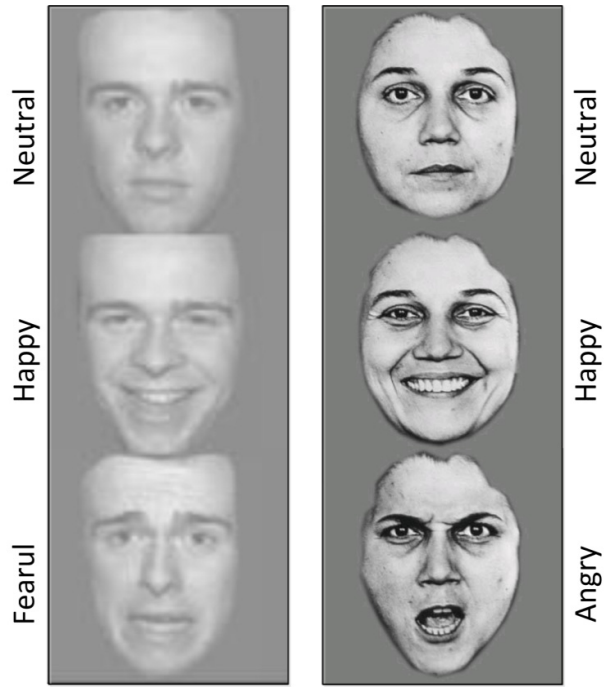

Figure 1: Face stimuli for case I and II (left) and case III and IV (right).

For group I and II experiment was run in four blocks running for approximately five minutes. In each block 30 stimuli (10 stimuli of each facial expression) were presented at random order where each stimulus was repeated twice. For the groups III and IV the experiment had

180 trials (60 trials per emotion type) with stimuli presented in random order in two blocks of 90 trials.

## References

Association, A.P., 2000. *Diagnostic and statistical manual of mental disorders: DSM-IV-TR*, American Psychiatric Pub.

Tottenham, N. et al., 2009. The NimStim set of facial expressions: judgments from untrained research participants. *Psychiatry Research*, 168(3), pp.242–249.

Valla, J.-P., Bergeron, L. & Smolla, N., 2000. The Dominic-R: A pictorial interview for 6-to 11-year-old children. *Journal of the American Academy of Child & Adolescent Psychiatry*, 39(1), pp.85–93.

Young, A. et al., 2002. Facial Expressions of Emotion: Stimuli and Tests (FEEST) Thames Valley Test Company, Bury St. *Edmunds, UK.*